\documentclass[%
 reprint,
 amsmath,amssymb,
 aps,
]{revtex4-2}

\usepackage{graphicx}
\usepackage{dcolumn}
\usepackage{bm}
\usepackage{subfigure}
\usepackage{float}
\usepackage{amsthm,amsmath,amssymb}
\usepackage{mathrsfs}
\usepackage{color} 
\usepackage{acronym}
\usepackage{hyperref}
\usepackage{microtype}
\usepackage[strings]{underscore}

\hypersetup{colorlinks=true,
linkcolor=blue,
anchorcolor=blue,
citecolor=blue}

\newcommand{\TRC}{MOE Key Laboratory of TianQin Mission, %
TianQin Research Center for Gravitational Physics $\&$ School of Physics and Astronomy, %
Frontiers Science Center for TianQin, %
Gravitational Wave Research Center of CNSA, %
Sun Yat-sen University (Zhuhai Campus), %
Zhuhai 519082, China}

\newcommand{\HBA}{ Department of Physics, Hebei University, Baoding 071002, China}

\newcommand{\HBB}{ Hebei Key Laboratory of High-precision Computation and Application of Quantum Field Theory, Baoding 071002, China}

\newcommand{\HBC}{ Hebei Research Center of the Basic Discipline for Computational Physics, Baoding 071002, China}

\begin{document}

\preprint{APS/123-QED}

\title{Searching for gravitational-wave bursts with space-borne detectors }

\author{Zheng Wu}
 \affiliation{\TRC}
\author{Hui-Min Fan}
\email{fanhm3@mail.sysu.edu.cn}
 \affiliation{\HBA, and \HBB, and \HBC}
\author{Ik Siong Heng}
\affiliation{
 SUPA, School of Physics and Astronomy, University of Glasgow, Glasgow G12 8QQ, United Kingdom
}%

\author{Yi-Ming Hu}
\email{
corresponding author: huyiming@mail.sysu.edu.cn
}%
 \affiliation{\TRC}


\date{\today}

\begin{abstract}

The millihertz gravitational wave band is expected to be opened by space-borne detectors like TianQin. 
Various mechanisms can produce short outbursts of gravitational waves, whose actual waveform can be hard to model.
In order to identify such gravitational wave bursts and not to misclassify them as noise transients, we proposed a proof-of-principle power excess method, that utilized the signal-insensitive channel to veto noise transients.
We perform a test on simulated data, and for bursts with a signal-to-noise ratio of 20, even with the contamination of noise transient, our methods can reach a detection efficiency of 97.4\% under a false alarm rate of once per year. However, more frequent occurrences of noise transients would lower the detection efficiency.

\end{abstract}


\maketitle

\acrodef{GW}{gravitational wave}
\acrodef{CBC}{compact binary coalescence}
\acrodef{EMRB}{extreme-mass-ratio burst}
\acrodef{TDI}{time delay interferometry}
\acrodef{SNR}{signal-to-noise ratio}


\section{\label{sec:level1}INTRODUCTION}\label{Introduction}

During the first three observation runs of ground-based \ac{GW} detectors,
almost one hundred \ac{GW} signals from \acp{CBC} have been observed \cite{LIGOScientific:2018mvr,LIGOScientific:2020ibl,KAGRA:2021vkt}. 
In addition to search pipelines that are dedicated to \acp{CBC}, one can also use generic burst search pipelines to perform a model-independent search which looks for power excess in a limited time-frequency range.
The burst search pipelines are constructed to identify unexpected signals that are recorded \cite{LIGOScientific:2020ibl, LIGOScientific:2021tfm, LIGOScientific:2021izm, LIGOScientific:2021nrg, KAGRA:2021tnv, KAGRA:2021bhs, Klimenko:2015ypf, Drago:2020kic, Cornish:2020dwh, Cornish:2014kda}.
As a matter of fact, the first \ac{GW} event, GW150914, was initially identified through a burst search pipeline \cite{LIGOScientific:2016fbo}.

Space-borne \ac{GW} detectors like TianQin \cite{TianQin:2015yph, TianQin:2020hid} and LISA \cite{LISA:2017pwj} are expected to open the \ac{GW} window in the mHz frequency range. 
A large number of sources are expected to spend a long time in the band, emitting \ac{GW}s that can be used to study the underlying astronomy and physics \cite{Wang:2019ryf,Liu:2020eko,Huang:2020rjf,Fan:2020zhy,Liang:2021bde}.  
In addition to the long-lasting signals, a number of different mechanisms can also emit \ac{GW} signals that are short in duration and irregular in the waveform, like the cosmic string cusps and kinks \cite{Auclair:2023mhe,Auclair:2023brk},  memory effects \cite{Mitman:2020pbt}, supernova eruption \cite{Abdikamalov:2020jzn, Szczepanczyk:2023ihe}, and \acp{EMRB} \cite{Han:2020dql, Berry:2012im}. 

The existence of hard-to-model and even unexpected mechanisms calls for the need to develop waveform-independent burst search pipelines for space-borne \ac{GW} missions.
For the ground-based detectors, the biggest challenge for a burst search pipeline is to prove that the excess of power in a localized time-frequency range is astronomically originated, instead of noise transients \cite{Klimenko:2015ypf, Drago:2020kic, Cornish:2020dwh, Cornish:2014kda, Anderson:2000yy, Houba:2024tyn}.
The same challenge could also be applicable to space-borne missions, as noise transients have been identified in the LISA Pathfinder data \cite{Armano:2016bkm, Armano:2018kix, Baghi:2021tfd, LISAPathfinder:2022awx}.

The core principle for \ac{GW} burst search pipelines for ground-based detectors is the coincidence between multiple independent detectors \cite{Klimenko:2015ypf, Drago:2020kic, Cornish:2020dwh, Cornish:2014kda}. 
For space-borne \ac{GW} missions, a triangle detector like TianQin and LISA can be decomposed into three orthogonal channels.
Although the noise might not be independent, with proper construction, one can obtain the signal-insensitive channel or the \emph{null channel}\cite{Tinto:2001ui, Romano:2016dpx}.
If a burst is identified only in the observing channels, and not the null channel, then it is much more likely to be astronomically originated.
Otherwise, if the burst is obvious in both the observing channels and the null channel, then one should veto it as a noise transient.

In this work, we develop a proof-of-principle algorithm to distinguish \ac{GW} bursts from noise transients for future space-borne \ac{GW} missions. 
We discuss the implication of the TianQin observatory as an example, but the same principle can be easily transferred to similar missions like LISA. Both TianQin and LISA are expected to adopt \ac{TDI} to reduce laser frequency noise \cite{Tinto:2002de, Muratore:2021rwq}.
The commonly adopted combinations form two observing channels (A and E), and a null channel (T) \cite{Prince:2002hp}, which can be used to veto noise transients. Unlike previous work that uses the same class of waveforms for both bursts and noise transients\cite{Robson:2018jly}, we use astronomically motivated sources as simulated bursts and similar sine-Gaussian as noise transients.

This paper is organized as follows: Sec. \ref{Source} reviews the  \ac{GW} burst sources and the \ac{EMRB} waveform details used in the following analyses. 
Section \ref{TianQin} describes the response function, the \ac{TDI} data streams, and the simulated noise of the TianQin detector. 
Section \ref{data} discusses the insertion of signal and noise into the test data and the selection of adjustable parameters for noise transients. 
Section \ref{method} discusses the core methods of searching \ac{GW} bursts from simulated test data. 
Section \ref{Result} shows the whole process and results of the \ac{GW} burst search. 
Section \ref{Discussion} summarizes the method and discusses its effects and limitations. We work in natural units with $G=c=1$.

\section{SOURCES OF GRAVITATIONAL WAVE BURSTS}\label{Source}


A massive black hole binary can produce a burst of \ac{GW} signals when they approach a merger.
Other mechanisms, ranging from cosmic string cusps and kinks to EMRBs, can also produce short-duration \ac{GW} bursts.
During the operation of TianQin or LISA, if a nearby supernova explodes like SN 2023ixf did in M101 \cite{Yamanaka:2023gbr, Pledger:2023ick}, the burst monitoring capability would provide priceless information on the inner state of the supernova engine. 
Previous studies reveal that both TianQin and LISA can detect \ac{EMRB} signals \cite{Fan:2022wio, Berry:2012im}, which are short GW signals produced by the capture of compact objects (COs) by massive black holes (MBHs) in galaxies or dwarf galaxies, before evolving into extreme-mass-ratio inspirals. 
In this proof-of-principle study, we adopt the \ac{EMRB} waveform as injected burst signals.  

EMRB events are highly eccentric parabolic encounters of the CO with the MBH, with the majority of the energy released near the pericenter. These events are characterized by high mass ratio and high pericentric velocity, and their waveforms are usually obtained by approximate numerical kludge method\cite{Babak:2006uv, Berry:2012im, Fan:2022wio}. In this method, the CO is approximated as traveling along the geodesics in Kerr spacetime and radiates as in flat space-time.

The Kerr geodesics are parametrized by three quantities, which are power $\tilde{E}$, specific angular momentum along the symmetry axis $\tilde{L}_z$, and the Carter constant $\tilde{Q}$. For EMRBs, the COs are assumed to be subject to marginally bound orbits $\tilde{E}\sim1$ with a single pericenter passage. The evolution of the CO follows the three Kerr geodesic equations \cite{Berry:2012im}
\begin{equation}\label{geoEqus}
\begin{split}
(r^2+a^2\cos^2\theta)\frac{d\psi}{d\tau}=&\Big(2r_{\rm p}-(r_3+r_4)(1+\cos\psi)+\\
&\frac{r_3r_4}{2r_{\rm p}}(1+\cos\psi)^2\Big)^{1/2},\\
(r^2+a^2\cos^2\theta)\frac{d\chi}{d\tau}=&\sqrt{\tilde{Q}+\tilde{L}_z^2},\\
(r^2+a^2\cos^2\theta)\frac{d\phi}{d\tau}=&\Big(\frac{\tilde{L}_z}{\tilde{Q}\cos^2\chi/(\tilde{Q}+\tilde{L}_z^2)}-a+\\
&\frac{a}{\Delta}[(r^2+a^2)-\tilde{L}_za]\Big),\\
\end{split}
\end{equation}
where $\Delta=r^2+2Mr+a^2$, $M$ is the mass of MBH , $a$ is the dimensionless MBH spin, $\tau$ is the proper time, $r_p$ is the pericenter distance of the orbit, $r_3, r_4$ are the radial potential roots of the COs \cite{Babak:2006uv,Berry:2012im}.
$(\psi, \chi, \phi)$ are phaselike variable substitutes for spherical polar coordinates $(r, \theta, \phi)$ in order to overcome the numerical integration defects. 

The EMRB signals are short lived, and power $\tilde{E}$, angular momentum $\tilde{L}_z$, and the Carter constant $\tilde{Q}$ can be regarded as conserved. Integrating Eqs.(\ref{geoEqus}) then provides the trajectory of the COs.
 With the CO trajectory $\mathbf{x}(r, \theta, \phi)$, one can apply the quadrupole formula to derive the gravitational strain $h(t)$ \cite{Fan:2022wio}.

\section{THE TIANQIN OBSERVATORY}\label{TianQin}
The TianQin \ac{GW} observatory consists of three satellites with the Earth at the center of them. It aims to detect \ac{GW} sources in the frequency band $10^{-4}-1$\, Hz. It is located about $10^5$\, km away from the Earth, with arm lengths of about $1.7 \times 10^5$\, km. The norm of the constellation orbital plane points in the direction ($\beta={-4.7}^{\circ},\lambda = {120.5}^{\circ}$), that is, the sky position of the verification binary RX J0806.3 + 1527. Its Earth-orbiting period is about 3.6\, days. More detailed information about the orbit can be found in \citet[]{Fan:2020zhy} and \citet[]{Hu:2018yqb}. The nominal operation time of TianQin is five years, while a ``three months on, three months off" working scheme is assumed \cite{Fan:2020zhy}.

TianQin adopts laser interferometry to detect \ac{GW}s. 
When a GW passes the detector, it causes a variation in the arm length of the detector over time, which changes the laser phase shift. 
Assuming a propagating GW $h(\xi)$, its effect on the optical path length is
\begin{equation}\label{resSignal}
  \delta L_{ij}(t)=\frac{1}{2}\frac{\hat{r}_{ij}(t)\otimes\hat{r}_{ij}(t)}{1-\hat{n}\cdot\hat{r}_{ij}(t) } :\int^{\xi_j}_{\xi_i}h(\xi)d\xi,
\end{equation}
where $L_{ij}(t)$ is the instantaneous separation from satellite $i$ to satellite $j$, $\hat{r}_{ij}$ is the unit vector from satellite $i$ to satellite $j$, $\xi$ is the GW phase, and $``:"$ is the double contraction.

\subsection{Time delay interferometry}
Except for GW signals, instrumental noises like clock noise, tilt-to-length coupling noise\cite{Hartig:2022nxt, Hartig:2022htm, Chwalla:2020srv, Trobs:2017msu}, and laser noise \cite{Armano:2024hlv}will also cause a phase shift; the detailed budget can be found in \cite{TianQin:2015yph}. Among these instrumental noises, the laser frequency noise is the one of noise sources that should be suppressed. 
For ground-based detectors like LIGO and Virgo, the laser frequency noise experiences the same delays within the two equal-length arms and cancels, while for space-borne GW detectors, the arm lengths vary in time due to the movement of satellites. Therefore, an ideal equal-arm Michelson interferometer cannot be sustained.
The \ac{TDI} method can suppress the laser frequency noise by constructing an equal-arm Michelson interferometer through the appropriate combination of the data streams \cite{Tinto:2002de, Prince:2002hp}.

The TianQin constellation consists of three satellites that can form three TDI interferometers. Here, we apply the first-generation Michelson TDI\cite{Tinto:2022zmf, Tinto:1999yr, Baghi:2020ygw}. Let $i$ denotes the satellite at each vertex, where $i=1,2,3$. A specific TDI interferometer example by using satellite 1 is \cite{Rubbo:2003ap}

\begin{equation}\label{tdisignal}
\begin{aligned}
\mathrm{X}(t)=&\Phi_{13}(t-L_{21}-L_{12}-L_{31})+\Phi_{31}(t-L_{21}-L_{12})\\
&+\Phi_{12}(t-L_{21})+\Phi_{21}(t)\\
&-\Phi_{12}(t-L_{31}-L_{13}-L_{21})-\Phi_{21}(t-L_{31}-L_{13})\\
&-\Phi_{13}(t-L_{31})-\Phi_{31}(t).\\
\end{aligned}
\end{equation}

\begin{equation}\label{phase difference}
\begin{aligned}
\Phi_{ij}=&C_i(t_j-L_{ij})-C_j(t_j)\\
&+2\pi\nu_0[n_{ij}^p(t_j)-n_{ij}^a(t_j)+n_{ji}^a(t_j-L_{ij})+\delta L_{ij}]
\end{aligned}
\end{equation}
where $\Phi_{ij}$ is the phase difference measured on satellite $j$ compared with the phase from satellite $i$. $C_{i}$ is the laser frequency noise introduced on satellite $i$, $t_j$ is the time measure on satellite $j$, $\nu_0$ is the laser frequency, $n^p$ is the displacement noise and $n^a$ is the acceleration noise. The remaining laser frequency noise among this TDI data is
\begin{equation}\label{TDI}
\begin{aligned}
C_{\rm X}=&C_{1}(t-L_{21}-L_{12}-L_{31}-L_{13})\\
&-C_{1}(t-L_{31}-L_{13}-L_{21}-L_{12}).
\end{aligned}
\end{equation}

In reality, the arm length $L_{ij}$ varies over time and has different values, so $C_{\rm X}$ may not be zero \cite{TianQin:2015yph}, and higher generation of \ac{TDI} would be needed.
However, for simplicity, we assume that the arms of the interferometer are not equal but constant and that the laser frequency noise is completely canceled.

$\rm X$ in Eq. (\ref{tdisignal}) is measured in vertex $1$. By permutation of the indices, we can get similar forms for $\rm Y$ and $\rm Z$ in vertex $2$ and vertex $3$.


\subsection{Group of channels A E T}\label{AETChannel}


The interferometric data channels X, Y, and Z are not independent. They can be linearly combined to form an orthogonal channel group (assuming equal noise levels on each satellite) A, E, and T, which are given by:
\begin{eqnarray}\label{equ::AET}
\begin{split}
\mathrm{A}=&\frac{1}{\sqrt{2}}(\mathrm{Z}-\mathrm{X}),\\
\mathrm{E}=&\frac{1}{\sqrt{6}}(\mathrm{X}-2\mathrm{Y}+\mathrm{Z}),\\
\mathrm{T}=&\frac{1}{\sqrt{3}}(\mathrm{X}+\mathrm{Y}+\mathrm{Z}).
\end{split}
\end{eqnarray}
~\\
\noindent Among these channels, channel T which we call the ``Sagnac observable''\cite{Tinto:2002de, Prince:2002hp, Muratore:2021uqj, Hogan:2001jn}, is much less sensitive to GW signals below the characteristic frequency $f_*=1/2\pi L$, where $L$ is the arm length.
It is also known as the null channel\cite{Tinto:2002de, Prince:2002hp, Muratore:2021uqj, Romano:2016dpx}, or noise monitoring channel.

The null channel was first proposed to discriminate a gravitational wave background from instrumental noise in the LISA detector\cite{Hogan:2001jn, Tinto:2001ui, Romano:2016dpx}. \citet{Muratore:2021rwq} found that there are many combinations of null channel that exhibit different sensitivities to gravitational waves. \citet{Robson:2018jly} transplanted the idea of BayesWave, a ground-based gravitational wave search pipeline, to a space-based gravitational wave detector, and also mentioned the role of the null channel for detecting gravitational wave bursts.

\subsection{Noise simulation}\label{noise}
\begin{figure}
\centering
\includegraphics[width=0.5\textwidth,height=0.4\textwidth]{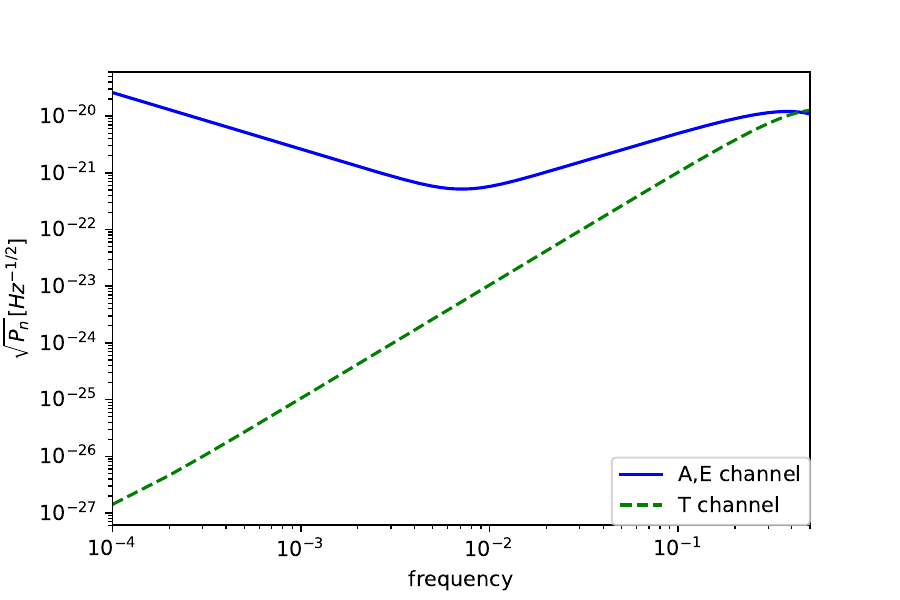}
\caption{The noise amplitude spectral density curve of A, E, and T channels. The noise amplitude spectral density of the A or E channel is significantly stronger than that of the T channel below the characteristic frequency($f_* \approx 0.28$ Hz for TianQin)} \label{figpsd}
\end{figure}

After suppressing the laser frequency noise by the TDI, the noise model of TianQin is characterized by $S_{\rm a}^{1/2}=1\times10^{-15}\rm m~s^{-2}/Hz^{1/2}$, which describes the residual acceleration on a test mass playing as an inertial reference, and $S_{\rm p}^{1/2}=1\times 10^{-12}\rm m/Hz^{1/2}$, which describes the displacement noise measured with the intersatellite laser interferometer. In the A, E, and T channels, the single-sided noise power spectral density of TianQin is \cite{Liang:2021bde}
\begin{widetext}
\begin{eqnarray}
P_{\mathrm{n}_{\mathrm{A / E}}}(f)=\frac{2 \sin ^{2}\left[\frac{f}{f_{*}}\right]}{L^{2}}\left[\left(\cos \left[\frac{f}{f_{*}}\right]+2\right) S_{\mathrm{p}}(f)+2\left(\cos \left[\frac{2 f}{f_{*}}\right]+2 \cos \left[\frac{f}{f_{*}}\right]+3\right) \frac{S_{\mathrm{a}}(f)}{(2 \pi f)^{4}}\right]\label{AE}
\end{eqnarray}

\begin{eqnarray}
P_{\mathrm{nT}}(f)=\frac{8\sin^{2}\left|\frac{f}{f_{*}}\right|\sin^{2}\left|\frac{f}{2f_{*}}\right|}{L^{2}}\left[S_{\mathrm{p}}(f)+4\sin^{2}\left[\frac{f}{2f_{*}}\right]\frac{S_{\mathrm{a}}(f)}{(2\pi f)^{4}}\right]\label{T}.
\end{eqnarray}
\end{widetext} 
An illustration of these noise spectrums is given in Fig. \ref{figpsd}. According to the noise power spectral density, we can generate the Gaussian noise in the A, E, and T channels.  

Ground-based \ac{GW} detectors suffer from disturbance of noise transients known as glitches. Glitches are also found in the LISA Pathfinder data\cite{LISAPathfinder:2022awx, Armano:2018kix, Baghi:2021tfd}; therefore it is reasonable to prepare for the existence of noise transients in future TianQin or LISA data. 

A common type of glitche found in LISA Pathfinder can be described by sine-Gaussian \cite{Baghi:2021tfd}. EMRB signals also have a similar sine-Gaussian shape. Therefore, we choose to simulate glitches using sine-Gaussian (SG) functions.

For the simulated glitches, we adopt the following expression
\begin{eqnarray}
g_{\rm SG}(t)=A\exp \left(-\frac{\left(t-t_{0}\right)^{2}}{\tau^{2}}\right) \sin \left(2 \pi f_{0}(t-t_{0})\right),\label{SG}
\end{eqnarray}
where $A$ is the amplitude of the glitch and $t_0$ and $f_0$ are the central time and central frequency of the glitch, respectively. The duration parameter is chosen as $\tau=0.6/f_0$, which makes the shape of simulated glitches mimic the glitches in the data of LISA Pathfinder\cite{Baghi:2021tfd}.

Although glitches can be caused by many origins, one can roughly classify them into two categories by the way they contaminate data \cite{Robson:2018jly}: through the optical path or the acceleration of test masses. 
Different classes of glitches will have different responses in the A, E, and T channels.  
A noise transient in the optical path length pointing from satellite $i$ to $j$ can be labeled as $\Phi_{ij}^{\rm op}(t_j)=g_{\rm SG}(t_j)$, and an acceleration noise transient associated with the proof mass on satellite $j$ that is referenced against satellite $i$ can be labeled as $\Phi_{ij}^{\rm ac}(t_j)=-g_{\rm SG}(t_j)$. The optical path and acceleration glitches in the X, Y, and Z channels can be constructed by injecting these two types of noise transients into the appropriate term in Eq. (\ref{tdisignal}). In the A, E, and T channels, the response for these two types of noise transient in the frequency domain is listed in Table \ref{TableII} in the Appendix.

\section{DATA CONSTRUCTION}\label{data}

We start simulating the TianQin data. We label the data using two categories, with signal and without signal: 
\begin{eqnarray}\label{equ::dataT}
d_{\alpha}(t)=\begin{cases}h_{\alpha}(t)+n_{\alpha}(t)\\n_{\alpha}(t)
\end{cases}
\end{eqnarray}
where $\alpha=\rm A, E, T$, represents different channels, $h(t)$ is the response EMRB waveform, $n(t)=w(t)+g(t)$ is the noise, where $w(t)$ is the stationary Gaussian noise, and $g(t)$ is the simulated glitches. 

The noise and signal in the A, E, and T channels are generated separately. 
For the detector Gaussian noise, it is generated according to the noise power spectral density of Eqs. (\ref{AE}) and (\ref{T}). 
For the transient noise, it is generated according to Eq. (\ref{SG}) and Table \ref{TableII}. 
The central frequency and amplitude of $g_{\rm SG}(t)$ are randomly drawn between $10^{-4}-0.5\rm\,Hz$ and  $10^{-22}-10^{-18}$ (with equal
probability across different orders). The central time $t_0$ of $g_{\rm SG}(t)$ is randomly drawn in the duration of each data stream. Based on results from the LISA Pathfinder satellite\cite{Baghi:2021tfd}, we consider two cases for the average time interval between noise transients, namely 10,000 and 100,000 seconds, respectively. 

\begin{figure}
\centering
\includegraphics[width=0.5\textwidth,height=0.4\textwidth]{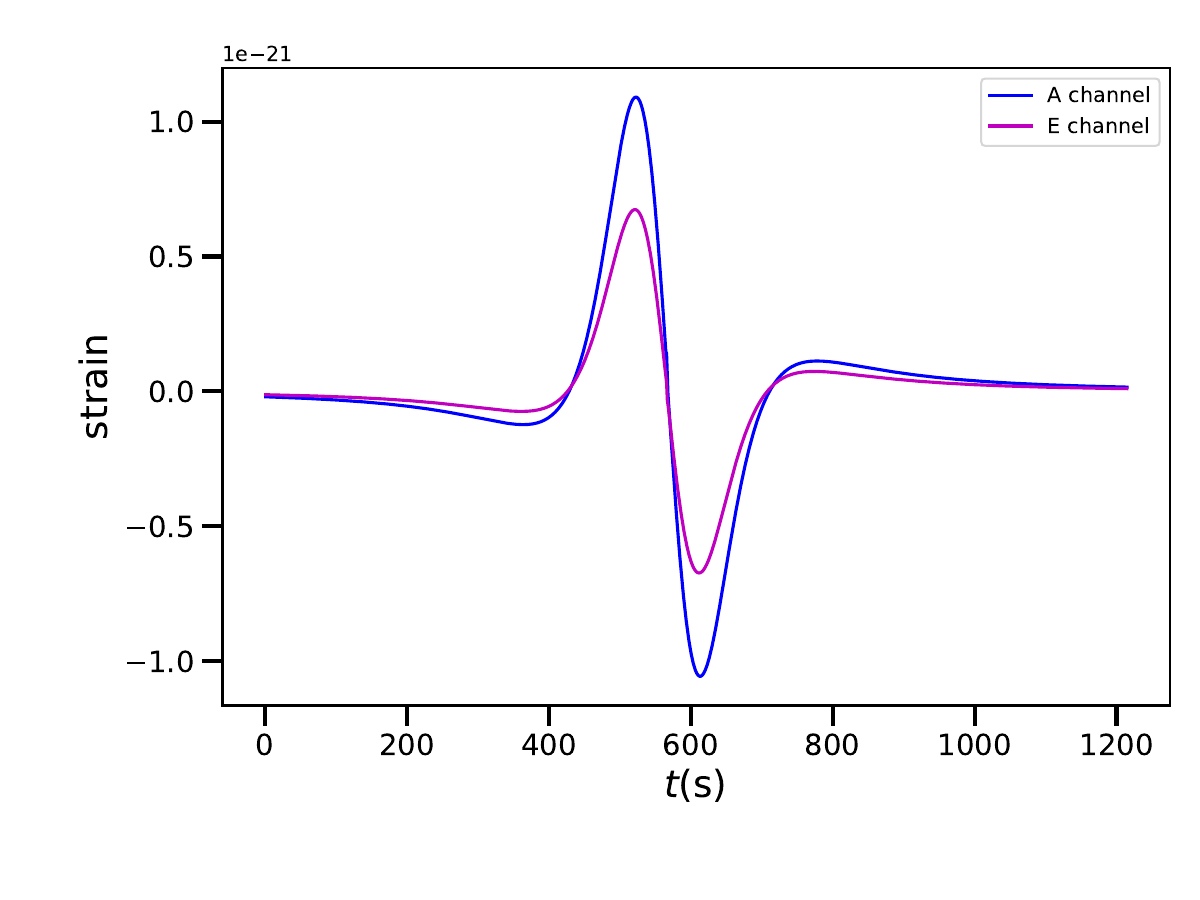}
\caption{An EMRB waveform of the A,E channels with $M=10^6\,M_\odot, r_p=6\,M, m=10\,M_\odot$, luminosity distance $D=1\,\rm Mpc$, and MBH spin $a=0.98$. It is similar in morphology to the sine-Gaussian function. } \label{EMRBwave}
\end{figure}

The parameters of waveforms are sampled according to the astronomical model introduced in \citet[]{Fan:2022wio}. A simple analytical expression relating the EMRB strain to the parameters can be found in \cite{Berry:2013poa}. For a typical EMRB source with $M=10^6 M_{\odot}$ and $m=10 M_{\odot}$ and $r_p=6 r_g$ ($r_g$ is the Schwarzschild radius), the waveform characteristic can be found in Fig. \ref{EMRBwave}. The waveform duration of the EMRB is related to $M$ and $r_p$. The larger $M$ and $r_p$, the longer the waveform duration. The duration is proportional to $M$; however, the relationship between $r_p$ and duration is more complicated (it is difficult to give an analytical formula).


We generate four types of data streams, each 100,000 seconds long. They can be divided into two pairs of data streams: Type I and Type II, and Type III and Type IV. Type I and Type III are noise only, while Type II and Type IV contain signals. For Type I and Type II, the glitch occurs at a rate of once per 100,000 seconds, and for Type III and Type IV, it's once per 10,000 seconds.



\section{METHOD}\label{method}

We aim to develop a proof-of-principle algorithm to confidently identify \ac{GW} burst signals against Gaussian noise and glitches. 
Our only assumption for a \ac{GW} burst is that it is short and concentrated in the frequency range.
The fundamental principle is to use the signal-insensitive T channel as the \emph{null channel} and to veto noise transients.

\begin{figure}
\centering
\includegraphics[width=0.5\textwidth,height=0.4\textwidth]{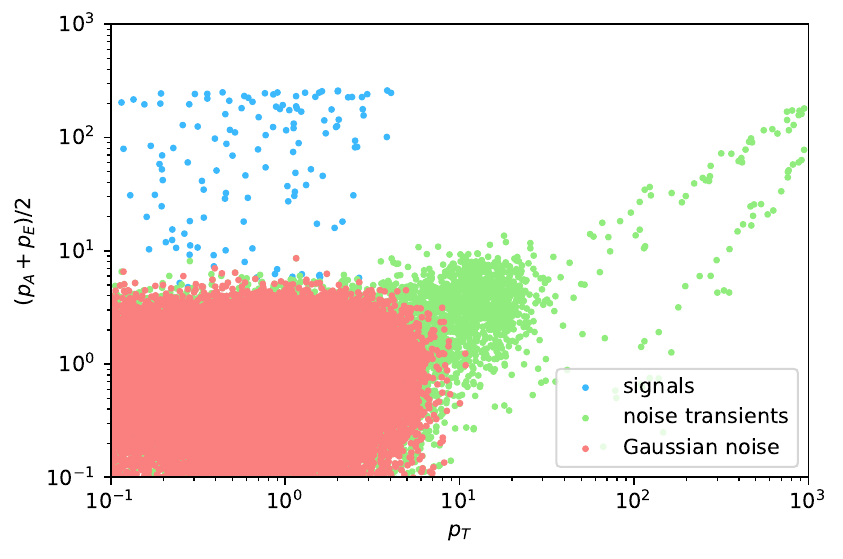}
\caption{The pixel power distribution of the Gaussian noise, the burst signal, and the noise transient in A, E, and T channels. Each pixel contains Gaussian noise. Blue dots represent pixels containing GW signals; green dots represent pixels containing noise transients; and red dots represent pixels containing only Gaussian noise. 
The green dots form two lines at the high values of $p_T$, representing optical path noise transients (lower) and acceleration noise transients (upper), respectively.
The example data streams consisted of Gaussian noise, one burst signal with an SNR of 112, and ten noise transients.} \label{fig0}
\end{figure}

Our method is mainly based on the characteristics of Gaussian noise, burst signal, and transient noise in the normalized spectrograms (if the data streams contain only Gaussian noise, the normalized power expectation for each pixel in the spectrogram is 1) of A, E, and T channels. To better illustrate the difference in their characteristics, we take three data streams lasting $100,000$ seconds in the A, E, and T channels as an example. The data streams consist of Gaussian noise, one burst signal with an SNR of 112, and ten noise transients. The time-frequency spectrograms with a frequency range between $10^{-4} and \sim 0.5$\,Hz consist of pixels with a time duration of $10,000$\,seconds and a frequency size of $10^{-4}$\,Hz.
This is determined by the designed lower frequency limit of $10^{-4}$\,Hz for TianQin.
Therefore, we have 50,000 pixels in each spectrogram. Since A and E channels are two orthogonal interferometers with similar responses to GW signals, we use $(p_{\rm A}+p_{\rm E})/2$ to represent the pixel normalized power of A and E channels, and $p_{\rm T}$ to represent the pixel normalized power of the T channel (power calculation details shown in Sec. \ref{candidate search}). We show the pixel normalized power in Fig. \ref{fig0}. In this figure, the blue dots represent pixels containing GW signals, the green dots represent pixels containing pure noise (noise transients and Gaussian noise ), and the red dots represent pixels containing only Gaussian noise. It is evident that the GW signal pixels exhibit high power in the A and E channels, but no obvious power excess compared with Gaussian noise in the T channel.
The noise transient pixels on the other hand have high power in all A, E, and T channels.

From Fig. \ref{fig0} one can notice that the signal power in a spectrogram is usually distributed over multiple pixels; and if the SNR of the signal is small, the difference in pixels normalized power between signal and noise will be much slighter, and makes the identification task much more challenging.
Therefore, we cluster neighboring pixels into one \emph{box}, and we make decisions based on the combined power from all pixels within the box. 

The method can be summarized as follows. Data streams from A, E, and T channels are whitened and then transformed into the time-frequency representation. This process generates the normalized spectrograms, which are analyzed in the subsequent steps. Based on the power difference of the normalized spectrograms among the three channels, three steps are taken to pick up \ac{GW} bursts. First, divide the spectrogram into several nonoverlapping boxes with $o\times l$ pixels, where $o$ is the number of pixels along the time axis and $l$ is the number of pixels along the frequency axis, then sum the power over pixels within it to obtain the power of each box. Second, construct the detection statistic of the box based on the power within the A, E, and T channels. The ideal detection statistic should have a good ability to distinguish bursts from either Gaussian noises or glitches. Finally, the box with the detection statistic exceeding a predetermined threshold will be identified as containing a \ac{GW} burst. 


\subsection{The time-frequency analysis}\label{candidate search}

The traditional signal-to-noise ratio (SNR) in the frequency domain is defined as
\begin{eqnarray}
\rho^2=\sum_{k=1}^N\frac{2h_k^2}{\sigma_{n_k}^2},
\end{eqnarray}
where $h_k$ is the Fourier amplitude of the GW signal, $\sigma^2_{n_k}=0.5P_n(f)/(\text{d}t^2\text{d}f)$ is the expected variance of the noise component $n_k$ at the frequency bin $k$, $P_n(f)$ is the strain spectral density of the noise, $N$ is the number of Fourier frequency bins, $\text{d}f$ is the bin width, and $\text{d}t$ is the sampling interval of the time series.

There are both noise and signal in our data streams. Therefore, inspired by the SNR definition, we define the normalized power of the pixel in the $i^{\rm th}$ time segment and the $k^{\rm th}$ frequency bin as \cite{Wen:2005xn}:
\begin{eqnarray}
p_{\alpha}(i, k)=\left[\frac{2\lvert d_{k}^{i}\rvert^{2}}{\sigma_{n_{k}}^{2}}\right]=\left[\frac{2\lvert\left(h_{k}^{i}+n_{k}^{i}\right)\rvert^{2}}{\sigma_{n_{k}}^{2}}\right],
\end{eqnarray}
where $d$ is the data stream; $n$ is the noise; and $\alpha=\rm A, E, T$ stands for the different channels.
Dividing the spectrogram into several identical boxes with the size of $o\times l$ pixels, the power in each box is
\begin{eqnarray}
\mathbb{E}_{\alpha}(i,k)=~\sum_{a=1}^{o}\sum_{b=1}^{l}p_{\alpha}(i+a,k+b).
\label{normalized power}
\end{eqnarray}

Additionally, we use $\mathbb{E}_{\rm signal}=(\mathbb{E}_{\rm A}+\mathbb{E}_{\rm E})/2$ to represent the combined box power of A and E channels.





\subsection{The receiver operating characteristic curve}\label{check statistic}

If the noise is purely Gaussian, we just need to choose the size of the box. That is because the above statistic $\mathbb{E}_{\rm signal}$ is sufficient to select \ac{GW} bursts. 
However, the potential contamination of noise transients makes it necessary to use the noise monitoring T channel to veto glitches. 
We, therefore, use both $\mathbb{E}_{\rm signal}$ and $\mathbb{E}_{\rm T}$ to construct the detection statistic. 
To optimize the choice of the box size and detection statistic, we use the receiver operating characteristic (ROC) method.

The ROC curve was plotted with pairs of the true positive rate (TPR) versus the false positive rate (FPR) for every possible threshold of models. TPR and FPR are percentages of data that contain events exceeding the threshold, which can be defined as
\begin{equation}
\begin{aligned}
{\rm TPR}=&\frac{\rm TP}{\rm {TP+FN}},\\
{\rm FPR}=&\frac{\rm FP}{\rm {FP+TN}},
\end{aligned}
\end{equation}
where TP (true positive) and TN (true negative) represent the number of correct identifications of the presence and absence of an event, and FP (false positive) and FN (false negative) represent the number of wrong identifications of the presence and absence of a signal. 
Since the TPR and the FPR are defined from different samples, two groups of data are needed: one noise-only group to calculate FPR, and one signal-containing group to compute TPR.
The ROC curve is a commonly used tool to quantify the distinguishing ability of a method.
When sliding the threshold, both TPR and FPR are changing. The better the method is, the more separated the two distributions are, and the closer the ROC curve can approach the top left corner. 



To optimize the construction of the detection statistic, we experiment through a variety of combinations of $\mathbb{E}_{\rm signal}$ and $\mathbb{E}_{\rm T}$. 
The combination with the largest area under curve(AUC) will be identified as the ideal detection statistic.

\subsection{Choice of the threshold}\label{Choice of the threshold}

The detection statistics threshold is the value used to distinguish whether the candidate event is a signal or noise. In general, a higher threshold will increase the miss alarm probability, while a lower threshold will increase the false alarm rate (FAR). We set the detection statistics threshold to satisfy the equation: 
\begin{eqnarray}
{\rm FAR}=\langle{m}_{\rm 1yr}\rangle=\langle{m}\rangle{N} =1 {\rm yr}^{-1}\label{equation of threshold},
\label{Equ:threshold}
\end{eqnarray}
which means that the expectation of the false alarm number $\langle{m}_{\rm 1yr}\rangle$ in one year is 1. A year's worth of data can be equally divided into $N$ segments. Furthermore, the expectation of the false alarm number for one data stream denoted as $\langle m \rangle$, has the value $1/N {\rm yr}^{-1}$.



\section{RESULTS}\label{Result}
To obtain false positive rates and detection rates, we apply the same criteria to search for signals in data with and without signals. We generate four types of data streams, each 100,000 seconds long. They can be divided into two pairs of data streams based on the glitch occurrence frequency: Type I and Type II, and Type III and Type IV. There are 1000 of each type. We use the first two types of data streams to determine the optimal box size and detection statistics. In a more challenging case, noise transients occur more frequently. So we compare the detection performance in the above two pairs of types to show the impact of glitch occurrence frequency on the detection of gravitational wave bursts.


\subsection{The optimal size of the box}\label{The optimal size of the box}



We use the data streams of Type I and Type II in this section. Each spectrogram is generated by setting the pixels with a time duration of 10,000 seconds and a frequency resolution of $10^{-4}$Hz. Then, each spectrogram is divided into nonoverlapping boxes with the size of $o\times l$ pixels. 
Our experiments show that 10,000 seconds is sufficient to cover the majority of the power from a burst signal, therefore we set the optimal value of $o$ as 1.
We then explore the performance of the box with values of $l$ = 50, 100, 200, 250, 500 which correspond to frequency sizes of 0.0050, 0.0100, 0.0200, 0.0250, and 0.0500Hz.

We use $R=\mathbb{E}_{\rm signal}/\mathbb{E}_{\rm T}$ as the test detection statistic. The box with the maximum value $R_{\rm max}$ for a data stream is designated as the candidate for ROC.  Then, the ROC curve determines the appropriate box size that can be obtained.
 
 
 %



\begin{figure}[htbp]
\centering
\includegraphics[width=0.5\textwidth,height=0.4\textwidth]{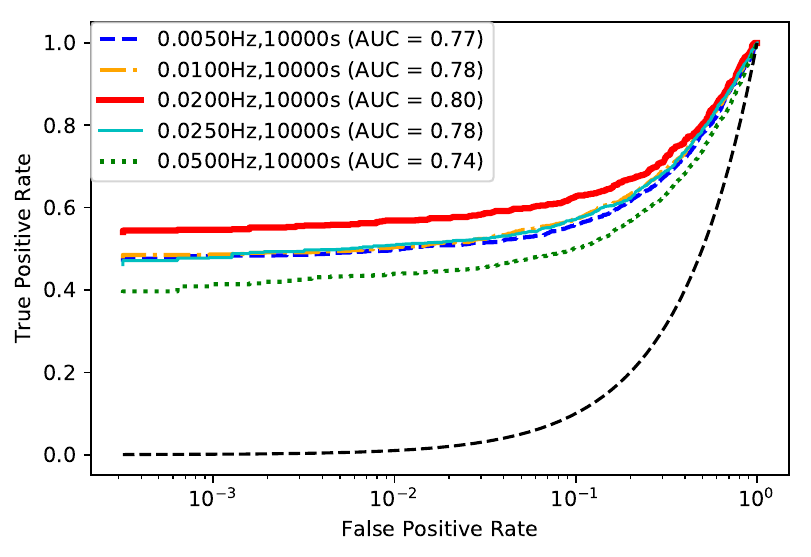}
\caption{The ROC curves for different box sizes of data streams. The curves in different colors correspond to boxes with frequency sizes of 0.0050, 0.0100 0.0200, 0.0250, and 0.0500 Hz, and a time duration of 10,000 seconds. The detection statistic is $\mathbb{E}_{\rm signal}/\mathbb{E}_{\rm T}$.  The dashed black line represents the diagonal line from (0,0) to (1,1).} \label{fig1}
\end{figure}



We present the results for different box sizes in Fig. \ref{fig1}. In this figure, the boxes with frequency sizes of 0.0050, 0.0100, 0.0200, 0.0250, and 0.0500 Hz correspond to AUC values of 0.77, 0.78, 0.80, 0.78, and 0.74, respectively. Therefore, we conclude that the suitable box has a frequency size of 0.02Hz corresponding to $l=200$, which can be explained by the fact that the full width at half maximum for the EMRB peaks at 0.02Hz. 

 

\begin{figure}
\centering
\includegraphics[width=0.5\textwidth,height=0.45\textwidth]{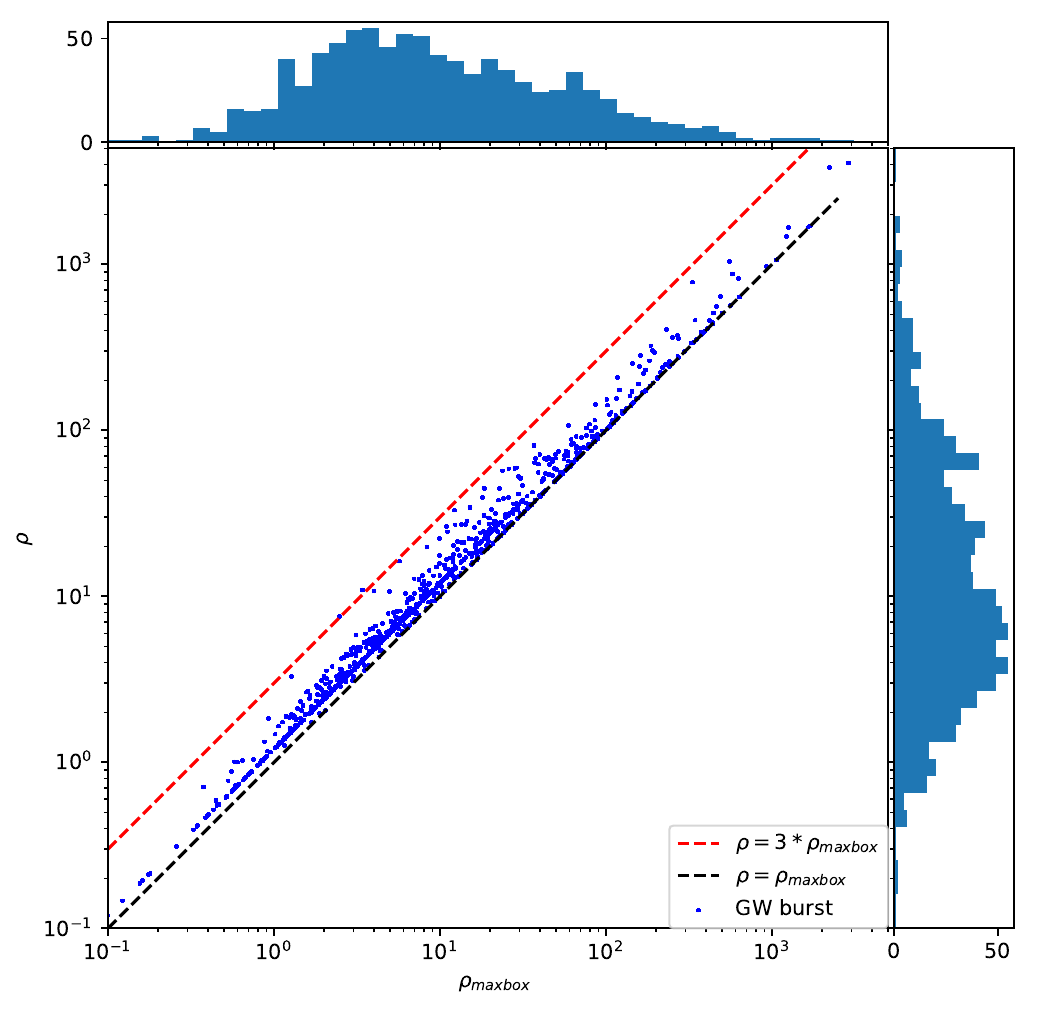}
\caption{The distribution of $\rho$ and $\rho_{\rm maxbox}$. Note that, for the simulated EMRB signals, the MBH mass is between  $M=10^4-10^9 M_{\odot}$, the CO mass is between $m=2.6-100 M_{\odot}$, and the luminosity distances ranges from 1 Mpc to 100 Mpc.} \label{figSNRrelation}
\end{figure}

The ROC results in Fig. \ref{fig1} do not show a high value of AUC. This is because most of our burst signals are too weak, which is a general characteristic of EMRB signals \cite{Fan:2022wio}. Even when we reduce the luminosity distance of the light source to 1\%, half of the signals still have an SNR of less than 10. However, this does not affect our conclusions as our goal is to find a better box size that can effectively gather signal power.
In Fig. \ref{figSNRrelation}, we present the SNR $\rho$ and the maximum SNR box components $\rho_{\rm maxbox}$ of those burst signals. From this figure, we can find that $\rho_{\rm maxbox}$ is approximately equal to $\rho$, which indicates that setting the box size to $1\times 200$ can effectively gather signal power.  



Note that, different signals may have different optimal box sizes. We just want to identify a suitable box size that works well for most of the burst signals. In the following sections, the box size with the value of $1\times 200$ is applied to the calculation of $R_{\rm max}$ for each data stream.


\subsection{The optimal expression of the detection statistic}\label{statistic optimal expression}
After we get the appropriate box size, the value of $\mathbb{E}_{\rm signal}$ and $\mathbb{E}_{\rm T}$ for each box is determined.
We then investigate the optimal construction of the detection statistic. We explore the different forms of $R(\mathbb{E}_{\rm signal}, \mathbb{E}_{\rm T})$, and use ROC to identify the optimal detection statistic that can efficiently distinguish burst signals from noise transients. Note that, we still use the data streams of Type I and Type II in this section.


\begin{table}[htbp]
  \begin{center}
   
    \begin{tabular}{l|c} 
      Forms & Area under ROC curve \\
      \hline
      $\sinh(\mathbb{E}_{\rm signal})/\sinh(\mathbb{E}_{\rm T})$ & 0.53 \\
      $\sinh(\mathbb{E}_{\rm signal})-\sinh(\mathbb{E}_{\rm T})$ & 0.51 \\
      $\exp(\mathbb{E}_{\rm signal})/\exp(\mathbb{E}_{\rm T})$ & 0.52  \\
      $\exp(\mathbb{E}_{\rm signal})-\exp(\mathbb{E}_{\rm T})$ & 0.50 \\
      $\sinh^{-1}(\mathbb{E}_{\rm signal})/\sinh^{-1}(\mathbb{E}_{\rm T})$&0.51  \\
      $\sinh^{-1}(\mathbb{E}_{\rm signal})-\sinh^{-1}(\mathbb{E}_{\rm T})$&0.50  \\
      $\log_{10}(\mathbb{E}_{\rm signal})/\log_{10}(\mathbb{E}_{\rm T})$ & 0.51  \\
      $\log_{10}(\mathbb{E}_{\rm signal})-\log_{10}(\mathbb{E}_{\rm T})$ & 0.50  \\
      $\mathbb{E}_{\rm signal}^\beta/\mathbb{E}_{\rm T}$ & 0.80 \\
      $\mathbb{E}_{\rm signal}^\beta-\mathbb{E}_{\rm T}$ & \quad0.52 \footnote{$\beta=1$} \\
    \end{tabular}
    \end{center}
  \caption{Area under ROC curve of different detection statistics.}
  \label{Tablestatistic}
\end{table}

The results are shown in Table \ref{Tablestatistic}. In this table, the left column is the detection statistic expressions, and the right column is the result of ROC. From this result, we can find that $\mathbb{E}_{\rm signal}^\alpha/\mathbb{E}_{\rm T}$ is the preferred detection statistic.


\begin{figure}
\includegraphics[width=0.5\textwidth,height=0.4\textwidth]{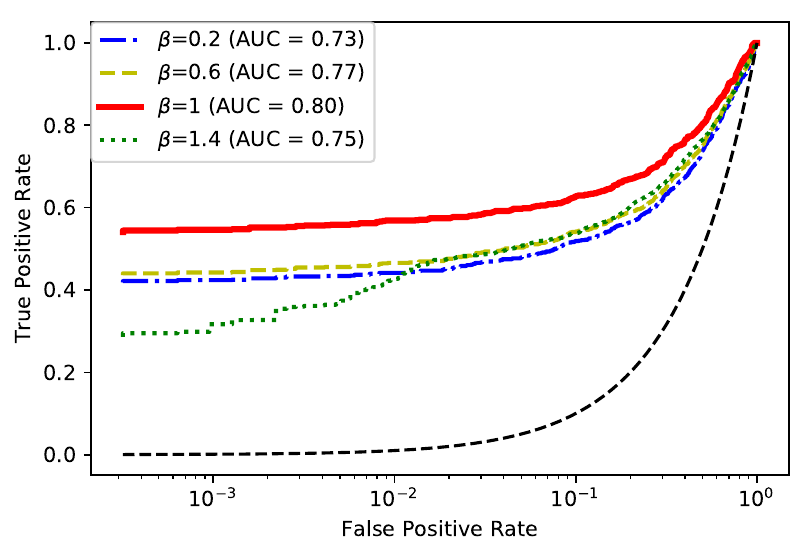}
\centering
\caption{The ROC curves of the detection statistic $\mathbb{E}_{\rm signal}^\beta/\mathbb{E}_{\rm T}$ computed from data streams with EMRB signal strength distribution shown in Fig. \ref{figSNRrelation}. The blue, yellow, red, and green dashed lines represent the power index $\beta$ with values of 0.2, 0.6, 1.0, and 1.4.  The dashed black line represents the diagonal line from (0,0) to (1,1).}
\label{fig2}
\end{figure}

We further examine the impact of the index $\beta$ on the ROC result and illustrate the finding in Fig. \ref{fig2}. In this figure, the blue,
yellow, red, and green dashed lines represent the ROC results of $\beta=0.2,\;\beta=0.6,\;\beta=1$, and $\beta=1.4$, respectively. The results indicated that $\beta=1$ is the optimal value for the detection statistic expression, which corresponds to  $\mathbb{E}_{\rm signal}/\mathbb{E}_{\rm T}$.

\subsection{The impact of glitch occurrence frequency}

After optimizing the box size and detection statistics above, we hope to evaluate the impact of glitch occurrence frequency on the detection of gravitational wave bursts in this section. We defined glitch occurrence frequency as the average time interval between noise transients. We evaluated the impact of glitch occurrence frequency by comparing the detection effect in two pairs of data streams with the average time interval between noise transients of 100,000(Type I and Type II) and 10,000 seconds(Type III and Type IV).

\begin{figure*}[htbp]
\centering

\subfigure[]{
\begin{minipage}[t]{0.5\textwidth}
\centering
\includegraphics[width=1\textwidth]{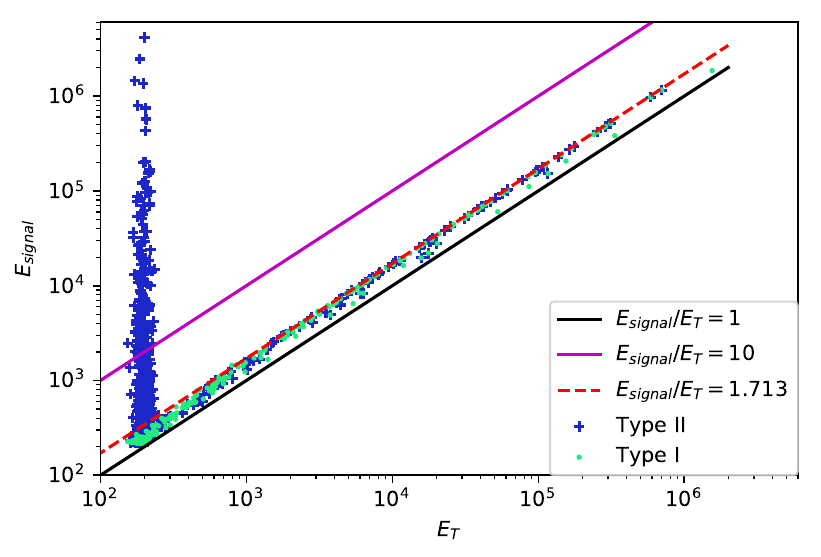}
\end{minipage}%
}%
\subfigure[]{
\begin{minipage}[t]{0.5\textwidth}
\centering
\includegraphics[width=1\textwidth]{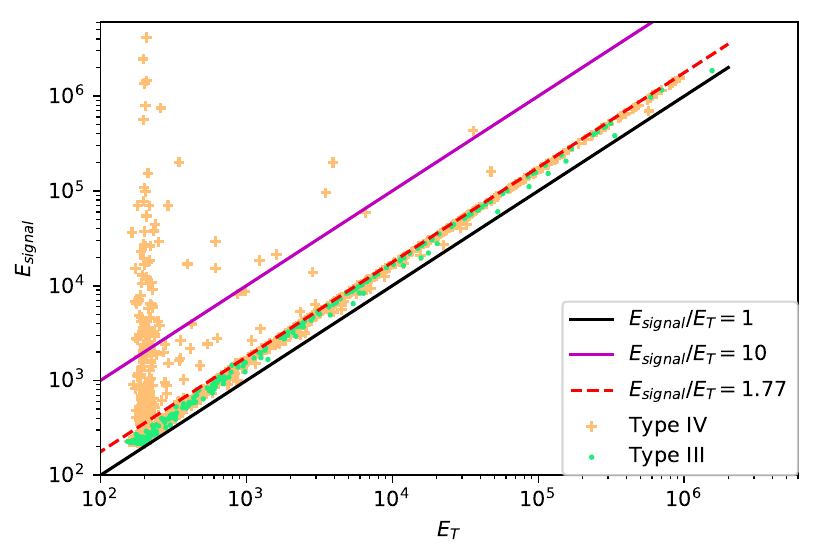}
\end{minipage}
}%

\centering
\caption{
Scatter plots of the power $\mathbb{E}_{\rm T}$ versus the power $\mathbb{E}_{\rm signal}$ for the boxes with the highest value of $R_{\rm max}$. The green dots and blue crosses in subplot (a) represent the boxes from the data streams of Type I and Type II, respectively. The green dots and orange crosses in subplot (b) represent the boxes from the data streams of Type III and Type IV, respectively. The burst signals are more distinguishable from the noise transients as $\rho_{\rm maxbox}$ increases. The solid lines show the different detection thresholds and the red dashed line shows the threshold corresponding to a false alarm rate of $1 {\rm yr}^{-1}$}
\label{distribution of different criteria}
\end{figure*}

The boxes with the maximum value of $R_{\rm max}$ for those four types of data streams are identified. We then mark their corresponding values of $\mathbb{E}_{\rm signal}$ and $\mathbb{E}_{\rm T}$ in Fig. \ref{distribution of different criteria}. In subplot (a) of Fig. \ref{distribution of different criteria}, the green dots correspond to the boxes selected from the data streams of Type I; the blue crosses correspond to the boxes selected from the data streams of Type II. In subplot(b), the green dots correspond to the boxes selected from the data streams of Type III; the orange crosses correspond to the boxes selected from the data streams of Type IV.

Figure \ref{distribution of different criteria} shows that our method can identify signals from data streams of Type II effectively. We applied a detection threshold of $R=1.713$ (we will explain the calculation process later) to distinguish the burst signals from the noise transients. However, data streams of Type III and IV have more noise transients, which pose challenges for signal detection. For example, in the spectrograms of Type IV, noise transients often coexist with signals in the same box, resulting in high values for both $\mathbb{E}_{\rm signal}$ and $\mathbb{E}_{\rm T}$. This can confuse the signal detection algorithm. Moreover, when a single noise transient has a certain probability of producing a false alarm, more frequent noise transients lead to more false alarms. Therefore, we need a higher detection threshold of $R$ to differentiate between data streams of Type III and IV.




\begin{figure*}[htbp]
\centering

\subfigure[]{
\begin{minipage}[t]{0.5\textwidth}
\centering
\includegraphics[width=1\textwidth]{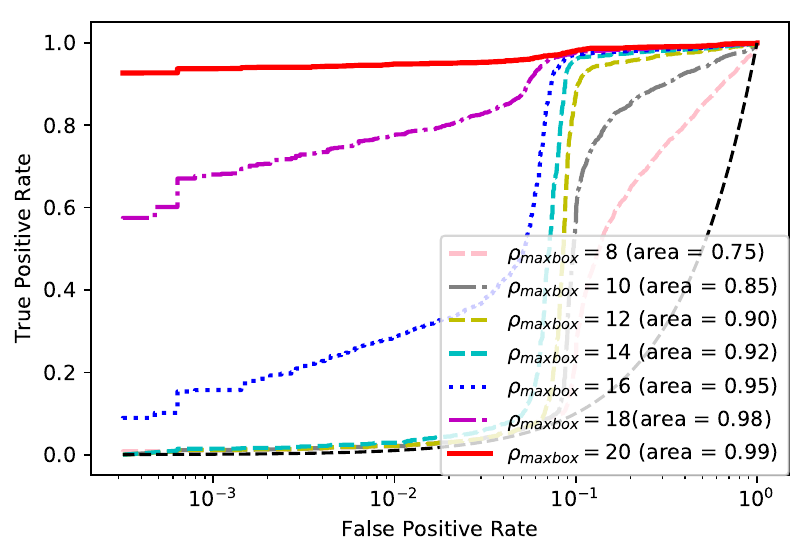}
\end{minipage}%
}%
\subfigure[]{
\begin{minipage}[t]{0.5\textwidth}
\centering
\includegraphics[width=1\textwidth]{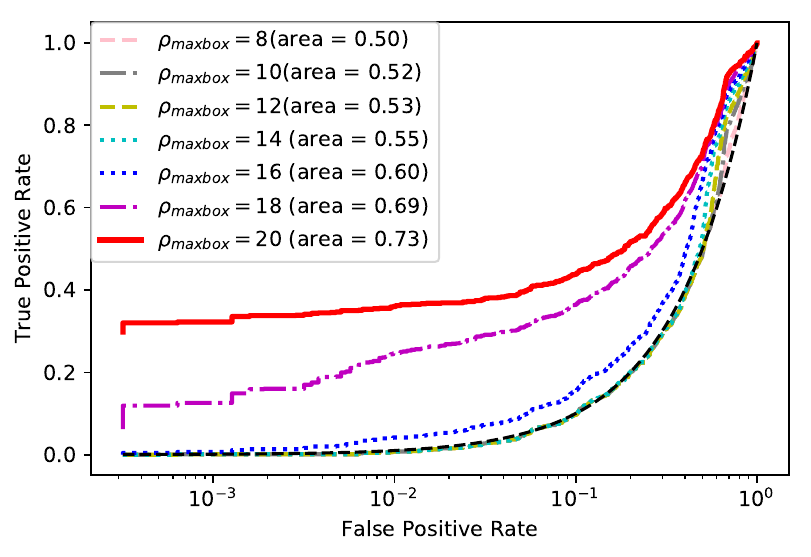}
\end{minipage}%
}%
\centering
\caption{The ROC curves for data streams with EMRB signals of different strengths ($\rho_{\rm maxbox}=8, 10, 12, 14, 16, 18, 20$). Subplot (a) shows the results for data streams of Type I and II, and subplot (b) shows the results for data streams of Type III and IV. The dashed black line is the diagonal line from (0,0) to (1,1). The curve in subplot (a) rises sharply near a false positive rate of 0.1 because of two types of noise transients that make the detection statistics of the noise data streams cluster around two distinct values. }
\label{ different SNR ROC}
\end{figure*}

We adjusted the luminosity distance of the burst signals and set their $\rho_{\rm maxbox}$ to fixed values: $\rho_{\rm maxbox}=8, 10, 12, 14, 16, 18, 20$. Figure \ref{ different SNR ROC} shows the ROC results, where subplot (a) is from data streams of Type I and II, and subplot (b) is from data streams of Type III and IV. This figure indicates that our method can identify signals with a strength $\rho_{\rm maxbox}>18$ from data streams of Type II effectively. However, signals from data streams of Type III need a much higher strength to separate them from noise transients. Another noteworthy point is that the distributions of $R_{\rm max}$ for type I and III data streams are quite different. The statistics of boxes containing optical path noise transients are significantly larger than those of boxes containing acceleration noise transients. If the average interval between noise transients is 10,000 seconds, each piece of the data stream has about 10 noise transients (optical path noise transients and acceleration noise transients occur with equal probability). Therefore, the distribution of $R_{\rm max}$ for noise type III data streams is mainly affected by optical path noise transients. If the average interval between noise transients is 100,000 seconds, each piece of the data stream has only about 1 noise transient. Therefore, the distribution of $R_{\rm max}$ for noise type I data streams is affected almost equally by both types of noise transients.



We calculated the detection threshold for the above two types of average intervals between noise transients using Eq. (\ref{Equ:threshold}). Each data stream has a length of 100,000 seconds, so there are $N$=315 independent data streams in one year. Therefore, the value of $\langle{m}\rangle$ for each data stream is 1/315. We found that the detection statistic threshold $R$ should be 1.713 for the average interval between noise transients of 100,000 seconds and 1.770 for the average interval between noise transients of 10,000 seconds



\subsection{The performance of the method}\label{The determination of the detection threshold}



\begin{figure}
\includegraphics[width=0.5\textwidth,height=0.4\textwidth]{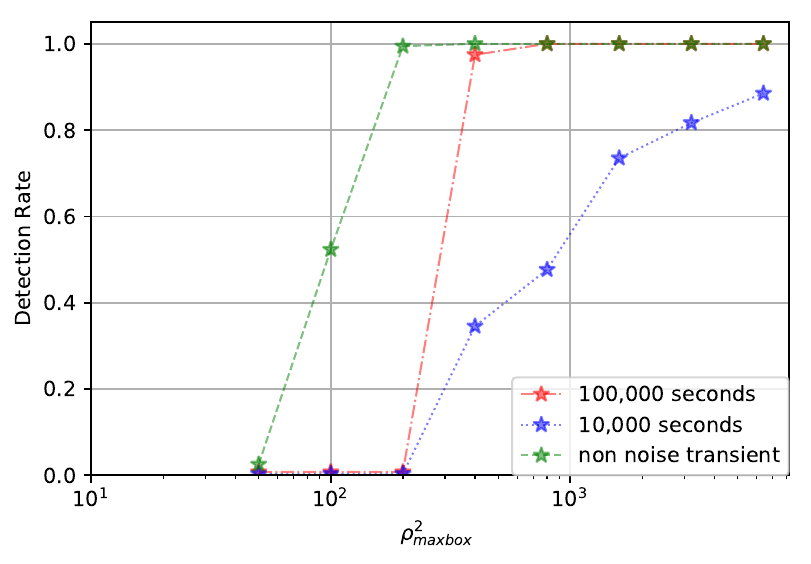}
\caption{The detection rate curve of signals varies with the values of $\rho_{\rm maxbox}$ in three different noise environments. The three curves correspond to three different noise environments: no noise transients (green), the average interval between noise transients is 100,000 seconds (orange), and the average interval between noise transients is 10,000 seconds (blue). The detection rate is 99.4\% for signals with $\rho_{\rm maxbox}=14.1$ in the absence of noise transients. For an average interval between noise transients of 100,000 seconds, the detection rate is 97.4\% for signals with $\rho_{\rm maxbox}=20.0$. For an average interval of 10,000 seconds, the detection rate drops to 88.5\% for signals with $\rho_{\rm maxbox}=80.0$.} \label{figAT}
\end{figure}

In this section, we present the performance of our method in three different noise environments: no noise transients, the average interval between noise transients is 10,000 seconds, and the average interval between noise transients is 100,000 seconds. We set the $\rho_{\rm maxbox}^{2}$ of the burst signals to eight fixed values that are uniformly distributed in log scale in the interval [50,6400]. Then, we evaluate the detection rate of the burst signals in each noise environment. Figure \ref{figAT} shows the results. The detection rate is 99.4\% for burst signals with $\rho_{\rm maxbox}=14.1 $ when there is no noise transient. The detection rate is 97.4\% for burst signals with $\rho_{\rm maxbox}=20.0 $ when the average interval between noise transients is 100,000 seconds. The detection rate drops to 88.5\% for burst signals with $\rho_{\rm maxbox}=80.0 $ when the average interval between noise transients is 10,000 seconds. Note that the performance of our method is not affected by the length of the data streams.

\section{DISCUSSION}\label{Discussion}
Noise transients, which can mimic GW burst signals, pose a challenge for searching generic GW bursts. In this work, we propose a proof-of-principle method to search GW burst signals for space-borne detectors. We construct three data streams called A, E, and T channels from the triangular configuration of the space-based GW detectors. Then, we use the ``null channel'' T, which is insensitive to GW signals, to distinguish transient noise from GW signals.


We test this method using mock data from the TianQin detector, a proposed space-based gravitational wave observatory. We inject EMRBs as the test signals and use the sine-Gaussian model to generate noise transients. We optimize the box size for capturing the signal power in the spectrogram of each data and find that a frequency width of 0.020 Hz and a time length of 10,000 seconds is a suitable box size. We also find that the optimal detection statistic is the ratio of the signal power to the noise power, $\mathbb{E}_{\rm signal}/\mathbb{E}_{\rm T}$. We set the detection thresholds based on a false alarm rate of $1 {\rm yr}^{-1}$ and find that they are 1.713 or 1.770 for an average noise transient interval of 10,000 seconds or 100,000 seconds, respectively. For an average noise transient interval of 100,000 seconds, we achieve a detection rate of more than 97.4\% for signals with $\rho_{\rm maxbox}>20$; for an average noise transient interval of 10,000 seconds, we achieve a detection rate of more than 88.5\% for signals with $\rho_{\rm maxbox}>80$. The results are not affected by the length of the data streams (if it is long enough to contain the entire signal).


Our method is tested based on the TianQin detector. However, it can be applied to the data processing of other space-borne GW detectors with triangular configurations. The main advantage of constructing the null stream in single detectors like TianQin is its independence from the propagation directions of GW signals and validity for multiple burst signals, which makes it a better choice compared with the null streams of the generic networks.

We assume that there is only one EMRB signal in each data sample, which makes it easier to apply the ROC method and check whether the power of the signal we find is consistent with that of the injected signal. In the future, one might relax this constraint and develop pipelines that can be sensitive to multiple bursts simultaneously.
We also aim to implement improvements in the future to make the method more robust. For example, we aim to give a custom box size and corresponding threshold to each candidate signal, go beyond the low-frequency limit for the signals, adopt higher generation TDI, include more realistic models for noises, etc.

\begin{acknowledgments}
The authors are grateful to Alejandro Torres Orjuela, Jian-Dong Zhang, Xue-Ting Zhang, and Chang-Qing Ye for their helpful discussions and suggestions.
This work has been supported by Guangdong Major Project of Basic and Applied Basic Research (Grant No. 2019B030302001), Hebei Natural Science Foundation (Grant No. A2023201041), and the Natural Science Foundation of China (Grants No. 12173104 and No. 12261131504). Ik Siong Heng is supported by the Science and Technologies Facilities Council (STFC Grant no. ST/V005634/1).
\end{acknowledgments}

\appendix

\section{Appendixes}\label{Appendixes}
We list the expression of TDI channels' response for noise transients linked to optical path and acceleration, respectively.
\begin{table*}[htbp]

  \begin{center}
    \begin{tabular}{l|c|c|c} 
      
      & $ \sqrt{2}\tilde{A} $ & $\sqrt{6}\tilde{E}$ & $\sqrt{3}\tilde{T}$ \\
      \hline
      $\Phi_{12}^{\rm op}$ & $-2i\tilde{g}_{\rm SG}e^{-2if/f_*}\sin{(f/f_*)}$ & $2i\tilde{g}_{\rm SG}e^{-if/f_*}\sin{(f/f_*)}(e^{-if/f_*}+2)$ &$2i\tilde{g}_{\rm SG}e^{-if/f_*}\sin{(f/f_*)}(e^{-if/f_*}-1)$\\
      $\Phi_{21}^{\rm op}$ & $-2i\tilde{g}_{\rm SG}e^{-if/f_*}\sin{(f/f_*)}$ & $2i\tilde{g}_{\rm SG}e^{-if/f_*}\sin{(f/f_*)}(2e^{-if/f_*}+1)$&$-2i\tilde{g}_{\rm SG}e^{-if/f_*}\sin{(f/f_*)}(e^{-if/f_*}-1)$\\
      $\Phi_{13}^{\rm op}$ & $-2i\tilde{g}_{\rm SG}e^{-2if/f_*}\sin{(f/f_*)}$& $-2i\tilde{g}_{\rm SG}e^{-if/f_*}\sin{(f/f_*)}(e^{-if/f_*}+2)$&$-2i\tilde{g}_{\rm SG}e^{-if/f_*}\sin{(f/f_*)}(e^{-if/f_*}-1)$\\
      $\Phi_{31}^{\rm op}$ &$2i\tilde{g}_{\rm SG}e^{-2if/f_*}\sin{(f/f_*)}$ & $2i\tilde{g}_{\rm SG}e^{-if/f_*}\sin{(f/f_*)}(e^{-if/f_*}+2)$ &$2i\tilde{g}_{\rm SG}e^{-if/f_*}\sin{(f/f_*)}(e^{-if/f_*}-1)$\\
      $\Phi_{23}^{\rm op}$ & $-2i\tilde{g}_{\rm SG}e^{-if/f_*}\sin{(f/f_*)}(e^{-if/f_*}+1)$ & $-2i\tilde{g}_{\rm SG}e^{-if/f_*}\sin{(f/f_*)}(e^{-if/f_*}-1)$& $-2i\tilde{g}_{\rm SG}e^{-if/f_*}\sin{(f/f_*)}(e^{-if/f_*}-1)$\\
      $\Phi_{32}^{\rm op}$ & $-2i\tilde{g}_{\rm SG}e^{-if/f_*}\sin{(f/f_*)}(e^{-if/f_*}+1)$ & $-2i\tilde{g}_{\rm SG}e^{-if/f_*}\sin{(f/f_*)}(e^{-if/f_*}-1)$ &$-2i\tilde{g}_{\rm SG}e^{-if/f_*}\sin{(f/f_*)}(e^{-if/f_*}-1)$\\
      $\Phi_{12}^{\rm ac}$ & 0 & $-8\tilde{g}_{\rm SG}e^{-if/f_*}\sin{(f/f_*)}^{2}$ &$4\tilde{g}_{\rm SG}e^{-if/f_*}\sin{(f/f_*)}^{2}$\\
      $\Phi_{21}^{\rm ac}$ & $4\tilde{g}_{\rm SG}e^{-if/f_*}\sin{(f/f_*)}^{2}$ & $-4\tilde{g}_{\rm SG}e^{-if/f_*}\sin{(f/f_*)}^{2}$ &$-4\tilde{g}_{\rm SG}e^{-if/f_*}\sin{(f/f_*)}^{2}$\\
      $\Phi_{13}^{\rm ac}$ & $-4\tilde{g}_{\rm SG}e^{-if/f_*}\sin{(f/f_*)}^{2}$ & $-4\tilde{g}_{\rm SG}e^{-if/f_*}\sin{(f/f_*)}^{2}$ &$-4\tilde{g}_{\rm SG}e^{-if/f_*}\sin{(f/f_*)}^{2}$\\
      $\Phi_{31}^{\rm ac}$ & $-4\tilde{g}_{\rm SG}e^{-if/f_*}\sin{(f/f_*)}^{2}$ & $4\tilde{g}_{\rm SG}e^{-if/f_*}\sin{(f/f_*)}^{2}$ & $4\tilde{g}_{\rm SG}e^{-if/f_*}\sin{(f/f_*)}^{2}$\\
      $\Phi_{23}^{\rm ac}$ & $4\tilde{g}_{\rm SG}e^{-if/f_*}\sin{(f/f_*)}^{2}$ & $4\tilde{g}_{\rm SG}e^{-if/f_*}\sin{(f/f_*)}^{2}$ & $4\tilde{g}_{\rm SG}e^{-if/f_*}\sin{(f/f_*)}^{2}$\\
      $\Phi_{32}^{\rm ac}$ & 0 & $8\tilde{g}_{\rm SG}e^{-if/f_*}\sin{(f/f_*)}^{2}$ &$-4\tilde{g}_{\rm SG}e^{-if/f_*}\sin{(f/f_*)}^{2}$\\
      
    \end{tabular}
    \caption{The response for optical path and acceleration noise transients in A, E, and T channels. For acceleration noise transients, the response in signal-sensitive channels (A and E channels) and the noise-only channel( T channel) are similar. For optical path noise transients,  the response in signal-sensitive channels (A and E channels) and the noise-only channel(T channel) are significantly different.}
  \label{TableII}
  \end{center}
\end{table*}


\bibliography{apssamp}


\end{document}